\documentclass[]{ptptex}

\usepackage{amsmath,amsfonts,amsbsy,bm}
\usepackage{graphicx}
\usepackage{slashed}
\usepackage{feynmf}

\def \pd {\partial}

\def \l {\mathcal{L}}

\def \ve {\varepsilon_n}
\def \vb {\varrho}
 
\def \even {\mathcal{E}}

\def \d3k {\frac{d^3k}{(2\pi)^3}}
\def \v#1{{\bm #1}}

\def \be {\begin{equation}}
\def \ee {\end{equation}}
\newcommand{\bal}{\begin{align}}
\newcommand{\eal}{\end{align}}
\newcommand{\bml}{\begin{multline}}
\newcommand{\eml}{\end{multline}}

\title{Quantum Electrodynamics in a Uniform Magnetic Field}
\author{Jun \textsc{Suzuki}\thanks{Present address: Department of Physics, National University of Singapore, Singapore 117542, Singapore}}

\inst{Department of Physics and Astronomy, University of South Carolina, Columbia, SC 29208}

\abst{
A systematic formalism for quantum electrodynamics in a classical uniform magnetic field is discussed.  
The first order radiative correction to the ground state energy of an electron is calculated. 
This then leads to the anomalous magnetic moment of an electron without divergent integrals. 
Thorough analyses of this problem are given for the weak magnetic field limit. 
A new expression for the radiative correction to the ground state energy is obtained. 
This contains only one integral with an additional summation with respect to each Landau level.  
The importance of this formalism is also addressed in order to deal with 
quantum electrodynamics in an intense external field.
}

\begin{document}

\maketitle


\section{Introduction} 
The discovery of the Dirac equation was essential to explain atomic phenomena.  
However, it turned out that nature requires modifications of the Dirac equation in several situations. 
The first difficulty was in explaining the phenomenon called ``Lamb shift." 
According to the Dirac equation, the energy levels $2S_{1/2}$ and $2P_{1/2}$ of the hydrogen atom are degenerate. 
However, experiment shows that these states are separated by 1057Mhz. This experimental fact cannot be 
explained purely from the Dirac equation in a Coulomb field \cite{itzykson}. 
There were many attempts to resolve this difficulty based on phenomenological arguments, 
but none of them could explain from the first principle. 
It was only after the establishment of quantum field theory that a satisfactory explanation to 
this problem was given.  Thus the quantum field theory of interactions between electrons and photons, 
quantum electrodynamics, was born and has been followed by great sequent successes. 

Quantum electrodynamics is one of the most successful idealized theories in physics. 
Agreements between theories and experiments are getting closer and closer with great precision. 
For example, the gyromagnetic factor of the electron, which is 2 according to the Dirac equation, 
is evaluated by one Feynman diagram in 1st order perturbation, 5 diagrams in 2nd order, 72 diagrams in 3rd order, 
and 891 diagrams in 4th order. Every higher order correction result justifies the precision test for quantum electrodynamics. 
There is no doubt this theory is correct. However all these results 
and all other computations higher than the second order perturbation involve the following mathematical difficulties. 
If one computes physical quantities, such as energy of the electron, the polarization of photons, etc., 
one encounters integrals which diverge typically in the ultraviolet region.  
Physicists are more optimistic. 
We solved this infinity problem with the help of renormalization techniques. 
Renormalization can extract finite results from infinities, which agree with experimental numbers. 
Although there is no logical reason and justification to adopt this working rule, 
quantum field theory cannot be applied without renormalization.  
Moreover, we need the remormalization procedure even in low energy physics, such as condensed matter physics. 

We should be satisfied with the present success of quantum field theory to some extent, 
and go further to solve more problems in natural phenomena. 
We should however, sometimes look back and ask ourselves fundamental questions. 

Dirac was probably the most well known physicist with pure soul. 
He insisted that this renormalization be considered as a temporary developing stage of theoretical physics. 
His attitude toward this problem can be seen from his works, for instance ``Quantum Electrodynamics without Dead Wood" \cite{dirac64}, 
and his last article ``The Inadequacies of Quantum Field Theory" \cite{dirac84}.
Unfortunately, Dirac could not solve this problem completely.  

In this paper we would like to revisit this problem from a different point of view. 
One of the key points in this problem is to recognize that we are dealing with infinite 
dimensional Hilbert space in quantum field theory. 
Associated with an infinite degrees of freedom, representations of the canonical (anti-)commutation relations 
cannot be uniquely determined. 
It was however, rarely used to explain physical phenomena until the work of Umezawa \cite{umezawa1,umezawa2}. 
His simple but elegant observation concluded that each different representation 
corresponds to different phases of physical systems.  
In a superconductor for example, superconducting states occur by a transition from 
one representation to another representation accompanying spontaneous breakdown of the gauge symmetry. 
Here one representation corresponds to normal conducting states and the other superconducting states. 
Thus the same Hamiltonian describing electrons in matter can describe different physical states. 
Therefore it is crucial to choose the proper representation of the canonical (anti-)commutation relations concerned, 
or in other words, the proper choice of the Fock space.  Since the Fock space is constructed by a cyclic operation of 
creation operators on the Fock vacuum, the correct Fock vacuum has to be chosen from infinite many vacua. 

From Umezawa's point of view, renormalization is understood as a procedure of 
changing the representation of operators. Since we have to stick to one 
representation throughout the calculations, a wrong initial choice of Fock space 
may result in a renormalization procedure. 
To be specific, if one starts from a non-interacting particle Fock space, it is necessary to 
carry out the renormalization procedure in order to calculate physical quantities for interacting particle systems. 
This observation seems to imply that any perturbation in quantum field theories require renormalization. 

With these observations it is quiet natural to apply the above idea to quantum electrodynamics. 
We will particularly consider the case where an electron is interacting with photons within a classical uniform magnetic field. 
The Lagrangian of the system is written as 
\be
\l_{tot}=\l_{el}+\l_{EM}+\l_{int},
\ee
where $\l_{el}$, $\l_{EM}$, and $\l_{int}$ are the Lagrangian for bare electrons, bare photons, and the interaction between 
electrons and photons respectively. In the presence of classical fields, the photon term is written 
\be
\l_{EM}=\l^{qm}_{EM}+\l^{cl}_{EM},
\ee
where $\l^{cl}_{EM}$ represents the Lagrangian for the given c-number classical fields. 
Hence, 
\be
\l_{tot}=\l_{el}+\l^{qm}_{EM}+\l^{cl}_{EM}+\l^{qm}_{int}+\l^{cl}_{int}.
\ee
The basic idea is to treat the first term and the last term together 
and quantize the electron field in a background of classical fields. 
Namely, we will consider 
\be
\l_{tot}=\tilde{\l}_{el}+\l^{qm}_{EM}+\tilde{\l}^{qm}_{int}, 
\ee
where $\tilde{\l}_{el}$ is the Lagrangian of electrons in classical fields, and 
$\tilde{\l}^{qm}_{int}$ is the interaction between electrons in classical fields and photons \cite{comment1}. 

A computation of the anomalous magnetic moment of the electron using 
this idea was, to our knowledge, first done by Luttinger in 1949 \cite{luttinger}. 
His observation was different from us, 
but he obtained the same first order perturbation result as is found with other methods but without divergent integrals. 
It seems that his work got little attention at the time. The main reason seems that his calculation was not 
relativistically covariant. Therefore, it was difficult to generalized to other processes. 
Later in his formulation of the proper time method, Schwinger independently obtained the same 
result \cite{schwinger51pr}. Based on Schiwinger's proper time method, Schwinger's collaborators, 
and Ritus and his collaborators extensively studied the above idea, 
i.e. quantizing the Dirac field in classical background fields \cite{comment2}. 
Their main interest is in applications to quantum electrodynamics in intense external fields, 
which cannot be handled in ordinary quantum electrodynamics. 

The main objective in this paper is to confirm Luttinger's work from the modern language, i.e. diagram techniques, 
and to establish the basic formulation in this problem.  We would also like to discuss the relation between his result and 
others' one. By reexamining this problem, we give an alternative expression for the first order 
radiative correction to the ground state energy. The final expression contains only one integral and a summation 
over the Landau levels. This new expression may be more useful than the previous result particularly for numerical evaluation. 
The physical interpretation is also clear by expressing each Landau level contribution separately.  
Another motivation to revisit this classical problem is from recent astrophysical observations of extremely strong magnetic field, 
such as magnetars and pulsars. 
The strength of magnetic fields in these objects are much above the critical field in quantum electrodynamics, 
i.e. $B_c=m^2c^3/(|e|\hbar)\simeq 4.4\times10^{13} G$.  
As will be discussed in this paper, the formalism present is essential in order to deal with strong external magnetic field problems. 

The organization of this paper is as follows.  In order to make our paper self-contained, first few sections are 
devoted to give a review on quantization of the Dirac field with classical c-fields. 
The readers who are familiar with this can skip sections 2$\sim$4 without any problems. 
Exact solution of the Dirac equation in a uniform magnetic field 
is reviewed in Sec.~2. Notations and conventions are also summarized in Sec.~2.  
The Dirac field is then quantized using this complete orthonormal basis in Sec.~3. 
The modified Dirac equation due to the radiative correction is discussed based on the mass operator formalism 
in Sec.~4.  One loop correction to the mass operator is evaluated in Sec.~5 and Sec.~6. 
The anomalous magnetic moment of an electron is calculated in a one loop level in Sec.~7. 
Divergent integrals appeared in this formalism is discussed in Sec.~8. 
In Sec.~9 a compact expression is given for the energy shift within the lowest Landau level approximation. 
The paper closes with conclusion and discussion in Sec.~10. 
Throughout the paper we set $\hbar=1$ and $c=1$. 

\section{Exact Solution of the Dirac equation} 
The exact solution of the Dirac equation in a uniform magnetic field is well known \cite{johnson}. 
To establish notations we give a brief review on the exact solution based on the operator formalism. 
The Dirac equation in an electromagnetic gauge field $A_{\mu}$ is
\be
(\slashed{\Pi}-m)|\psi\rangle=0,
\ee
where $\slashed{\Pi}=\gamma^{\mu}\Pi_{\mu}$ and $\Pi_{\mu}=P_{\mu}-eA_{\mu}$.  
The Dirac gamma matrices $\gamma^{\mu}$ satisfy the 
anti-commutation relation : 
\be
\{\gamma_{\mu},\gamma_{\nu}  \}=\gamma_{\mu}\gamma_{\nu}+\gamma_{\nu}\gamma_{\mu}=2g_{\mu\nu}I,
\ee
where $g_{\mu\nu}={\rm diag}(1-1-1-1)$ is the Minkowski metric, and $I$ is the $4\times4$ unit matrix. 
The electromagnetic field tensor $F_{\mu\nu}$ is defined by the commutation relation between $\Pi_{\mu}$ as 
\be \label{emtensor}
[\Pi_{\mu},\Pi_{\nu}]=-ieF_{\mu\nu}.
\ee
Here a charge of a particle is denoted by $e$ including its sign, for instance $e=-|e|<0$ for the electron. 
In eq. (\ref{emtensor}) the canonical commutation relation $[P_{\mu}, X_{\nu}]=ig_{\mu\nu}$ 
is used, and the gauge field $A_{\mu}$ is assumed as a function of $X_{\mu}$. 
(In this paper we express the position and momentum operators as $X^{\mu}$ and $P^{\mu}$ respectively. 
Their eigenvalues are denoted as $x^{\mu}=(t, x, y, z)$ and $p^{\mu}=(p_t, p_x, p_y, p_z)$.)
In the coordinate representation, the momentum operator is expressed as $P_{\mu}=i\pd/\pd x^{\mu}$. 
Then, the electromagnetic field tensor $F_{\mu\nu}$ reads a familiar form :  
\be 
F_{\mu\nu}=\pd_{\mu}A_{\nu}-\pd_{\nu}A_{\mu}.
\ee
By definition $F_{\mu\nu}$ is anti-symmetric, i.e. $F_{\mu\nu}=-F_{\nu\mu}$. 

Consider a uniform magnetic field where $B^i=(0,0,B)$ along the $z$-axis with $B$ a constant. 
The non-vanishing components of the field tensor $F_{\mu\nu}$ are then 
\be
F_{21}=-F_{12}=B, 
\ee
and all other components are zero. 
To solve the first order (in $P_{\mu}$) Dirac equation in external fields, 
it is convenient to bring it into the second order equation from 
multiplying by $\slashed{\Pi}+m$ from the left, 
\be \label{secondeq}
(\slashed{\Pi}+m)(\slashed{\Pi}-m)|\psi\rangle 
=(\Pi^{\mu}\Pi_{\mu}-m^2-\frac{e}{2}\sigma^{\mu\nu}F_{\mu\nu})|\psi\rangle.
\ee
Here $\sigma^{\mu\nu}=i[\gamma^{\mu},\gamma^{\nu}]/2$ and eq.~(\ref{emtensor}) is used. 
For the uniform magnetic field, the last term in eq.~(\ref{secondeq}) is 
\be
\frac{e}{2}\sigma^{\mu\nu}F_{\mu\nu}=-eB \Sigma^3=-eB 
\begin{pmatrix}
\sigma_z&0\\
0&\sigma_z\\
\end{pmatrix}
.
\ee
Therefore it is possible to decompose the Dirac four spinor into each component which is also 
an eigenspinor of the $z$-component of the spin matrix $\Sigma^3$.
Define the orthonormal spinors $w_a$ ($a=1,2,3,4$) by 
\be
w_1=
\begin{pmatrix}
1\\ 0 \\ 0 \\ 0
\end{pmatrix}
,\;
w_2=
\begin{pmatrix}
0\\ 1 \\ 0 \\ 0
\end{pmatrix}
,\;
w_3=
\begin{pmatrix}
0\\ 0 \\ 1 \\ 0
\end{pmatrix}
,\;
w_4=
\begin{pmatrix}
0\\ 0 \\ 0 \\ 1
\end{pmatrix}
.
\ee
These four spinors are eigenspinors of the spin matrix $\Sigma^3$ : $\Sigma^3w_a=\sigma_aw_a$ where 
eigenvalues $\sigma_a$ are 
\be
\sigma_a=+1\;(a=1,3),\quad \sigma_a=-1\;(a=2,4).
\ee
Expanding the four spinor by the eigenspinors $w_a$ as $|\psi\rangle=\sum_a w_a|\phi_a\rangle$, 
the second order equation for the four spinor is now reduced to the one for scalar functions $|f_a\rangle$ :
\be
(\Pi^{\mu}\Pi_{\mu}-m^2-\sigma_a\vb^2/2)|\phi_a\rangle=0. 
\ee
where $\vb=\sqrt{-2eB}$ with the assumption $-eB>0$, 
i.e. the direction of the constant magnetic field is upward along the $z$-axis. 
Since the only non-vanishing commutation relations among $\Pi_{\mu}$ is $[\Pi_1,\Pi_2]=ieB=-i\vb^2/2$, 
it is possible to reduce the problem to a two-dimensional one by separating these degrees of freedom. 
Choose the symmetric gauge for the gauge field : 
\be
A^{\mu}=(0,-B X^2/2,B X^1/2,0).
\ee
In this gauge $\Pi_0=P_0$ and $\Pi_3=P_3$. Then, the eigenfucntions for these operators are 
just plane waves with eigenvalues $p_t=E$ and $p_z$ respectively, e.g. $P_3|p_z\rangle=p_z|p_z\rangle$.  
Therefore, essentially the problem is to solve the following equation on the $xy$-plane : 
\be \label{2diff}
(\Pi_1^2+\Pi_2^2+m^2+\sigma_a\vb^2/2+p_z^2-E^2) |\phi'_a\rangle=0,  
\ee
where $|\phi_a\rangle=|E\rangle|\phi'_a\rangle|p_z\rangle$.

For later convenience, we solve eq.~(\ref{2diff}) with the aid of the complex variable representation 
(holomophic representation) for $\Pi_1$ and $\Pi_2$ \cite{itzykson87}. 
Introduce a complex variable $\xi=\vb(x+iy)/(2\sqrt{2})$ 
and its complex conjugate $\bar{\xi}=\vb(x-iy)/(2\sqrt{2})$ in the coordinate representation. 
The differential operators with respect to $\xi$ and $\bar{\xi}$ are related to $\pd_x=-iP_1$ and $\pd_y=-iP_2$ by 
\bal
\pd&\equiv \pd_{\xi}=\frac{\sqrt{2}}{\vb}(\pd_x-i\pd_y),\\
 \bar{\pd}&\equiv\pd_{\bar{\xi}}=\frac{\sqrt{2}}{\vb}(\pd_x+i\pd_y).
\end{align}
The covariant differential operators $D_1=-i\Pi_1$ and $D_2=-i\Pi_2$ are then expressed in terms of these 
new complex variables $\xi$, $\bar{\xi}$, $\pd$, and $\bar{\pd}$ : 
\bal
D_1&=\frac{\vb}{2\sqrt2}(\pd+\bar{\pd})- \frac{\vb}{2\sqrt2}(\xi-\bar{\xi}), \\
D_2&=i\frac{\vb}{2\sqrt2}(\pd-\bar{\pd})+i \frac{\vb}{2\sqrt2}(\xi+\bar{\xi}). 
\end{align}
Hence, 
\be 
\Pi_1^2+\Pi_2^2=-\frac 14 \vb^2(2\pd\bar{\pd}-2\xi\bar{\xi}-\pd\xi-\xi\pd+\bar{\pd}\bar{\xi}+\bar{\xi}\bar{\pd}).
\ee 
Finally, introduce two sets of creation and annihilation operators by  
\bal
\alpha^{\dagger}&=(-\bar{\pd}+\xi)/\sqrt{2},\ \alpha=(\pd+\bar{\xi})/\sqrt{2},\\
\beta^{\dagger}&=(-\pd+\bar{\xi})/\sqrt{2},\ \beta=(\bar{\pd}+\xi)/\sqrt{2}.
\end{align}
They satisfy the canonical commutation relations : 
\be
[\alpha,\alpha^{\dagger}]=[\beta,\beta^{\dagger}]=1,
\ee
and all other commutations are equal to zero.
It is easy to show 
\be
\Pi_1^2+\Pi_2^2=\vb^2(\alpha^{\dagger}\alpha+\frac 12), 
\ee
and this does not depend on the other set of creation and annihilation operators $\beta^{\dagger}$ and $\beta$. 
Then eq.~(\ref{2diff}) can be solved algebraically by 
\be
[\vb^2\alpha^{\dagger}\alpha+m^2+p_z^2+(\sigma_a+1)\vb^2/2-E^2] |\phi'_a\rangle=0.
\ee

Following an elementary exercise, the eigenvalues of $\alpha^{\dagger}\alpha$ are labeled by integers.
Since $\sigma_a=\pm1$, the energy eigenvalues are also labeled by integers $n$ : 
\be \label{energylevel}
E_n=\pm\sqrt{m^2+p_z^2+n\vb^2}\equiv\pm\ve\ (n=0,1,2,\cdots).
\ee
The ground state $|0\rangle$ of the system is defined by 
\be
\alpha|0\rangle=\beta|0\rangle=0.
\ee
The Hilbert space is then constructed by a cyclic operation of $\alpha^{\dagger}$ and $\beta^{\dagger}$ 
on $|0\rangle$, viz., 
\be
| n,\ell \rangle=\frac{(\alpha^{\dagger})^n(\beta^{\dagger})^{\ell}}{\sqrt{n!\ell!}} |0\rangle\ (n,\ell=0,1,2,\cdots).
\ee
These states consist of a complete orthonomal basis : 
\be \label{complete1}
\langle n,\ell|n',\ell'\rangle=\delta_{n,n'}\delta_{\ell,\ell'}. 
\ee
The completeness relation is satisfied,  
\be \label{complete2}
\sum_{n,\ell=0}^{\infty}|n,\ell\rangle\langle n,\ell |=\hat{I} ,
\ee
where $\hat I $ is the identity operator. 
Using this basis, the energy eigenstates corresponding to the energy $E_n$ are found as 
\bal 
|\phi'_a(n,\ell)\rangle&=| n-1,\ell \rangle\ (a=1,3), \\
|\phi'_a(n,\ell)\rangle&=| n,\ell \rangle\ (a=2,4), 
\end{align}
where the ground state ($n=0$) is realized only for $a=2,4$. 
To treat the ground state on the same footage as the other excited states, 
it is convenient to use the convention : 
\be
|-1,\ell\rangle\equiv 0\ {\rm for\ all}\ \ell .
\ee
Therefore the stationary solution of second order equation $(\Pi^{\mu}\Pi_{\mu}-m^2-\sigma_a\vb^2/2)|\phi_a\rangle=0$ is 
\bal 
|\phi_a(n,\ell,p_z;t)\rangle&=e^{-iE_nt} | n-1,\ell\rangle |p_z \rangle\ (a=1,3), \\
|\phi_a(n,\ell,p_z;t)\rangle&=e^{-iE_nt} | n,\ell\rangle |p_z \rangle\ (a=2,4). 
\end{align}
A similar convention will be used : 
\be
|\phi_a(-1,\ell)\rangle \equiv 0\ {\rm for\ all}\ \ell  .
\ee

The solution of  the original first order Dirac equation can be found as follows. 
Let $|\tilde{\psi}\rangle$ be the solution of second order Dirac equation : 
\be
(\slashed{\Pi}+m)(\slashed{\Pi}-m)|\tilde{\psi}\rangle=0, 
\ee
then $|\psi\rangle=(\slashed{\Pi}+m)|\tilde{\psi}\rangle$ is the solution of the first order Dirac equation : 
\be 
(\slashed{\Pi}-m)|\psi\rangle=(\slashed{\Pi}-m)(\slashed{\Pi}+m)|\tilde{\psi}\rangle=0.
\ee
This argument becomes empty when $|\tilde{\psi}\rangle$ is also a solution of the first order equation, 
i.e. $(\slashed{\Pi}-m)|\tilde{\psi}\rangle=0$.  It is straightforward to obtain an explicit representation 
of the operator $\slashed{\Pi}+m$ as follows :
\bal
\gamma^1\Pi_1+\gamma^2\Pi_2&=i\frac{\vb}{2}[\alpha(\gamma^1+i\gamma^2)-\alpha^{\dagger}(\gamma^1-i\gamma^2)]\\
&=i\vb
\begin{pmatrix}
0&0&0&\alpha\\
0&0&-\alpha^{\dagger}&0\\
0&-\alpha&0&0\\
\alpha^{\dagger}&0&0&0
\end{pmatrix}
,
\end{align}
and hence, 
\be \label{matrixop}
\slashed{\Pi}+m=
\begin{pmatrix}
m+P_0&0&P_3&i\vb\alpha\\
0&m+P_0&-i\vb\alpha^{\dagger}&-P_3\\
-P_3&-i\vb\alpha&m-P_0&0\\
i\vb\alpha^{\dagger}&P_3&0&m-P_0
\end{pmatrix}
.
\ee
The following standard representation of the gamma matrices are used in the above expressions :  
\be
\gamma^0=
\begin{pmatrix}
I&0\\
0&-I
\end{pmatrix}
,\quad 
\gamma^i=
\begin{pmatrix}
0&\sigma^i\\
-\sigma^i&0
\end{pmatrix}
,
\ee
where $I$ is the $2\times2$ unit matrix and $\sigma^i$ are the Pauli matrices. 
The positive energy solution is obtained by applying the operator $\slashed{\Pi}+m$ to $w_1|\phi_1\rangle$ and $w_2|\phi_2\rangle$ 
with $E_n=+\ve$,  
\bal \label{sol1}
|\psi_1(n,\ell,p_z;t)\rangle&=Ne^{-i\ve t}|p_z\rangle
\begin{pmatrix}
(\ve +m)|n-1,\ell\rangle\\
0\\
p_z|n-1,\ell\rangle\\
i\sqrt{n}\vb |n,\ell\rangle
\end{pmatrix}
,\\
|\psi_2(n,\ell,p_z;t)\rangle&=Ne^{-i\ve t}|p_z\rangle 
\begin{pmatrix}
0\\
(\ve +m)|n,\ell\rangle\\
-i\sqrt{n}\vb |n-1,\ell\rangle\\
-p_z|n-1,\ell\rangle
\end{pmatrix}
. \label{sol2}
\end{align}
The negative energy solution is obtained from $w_3|\phi_3\rangle$ and $w_4|\phi_4\rangle$ with $E_n=-\ve$, 
\bal \label{sol3}
|\psi_3(n,\ell,p_z;t)\rangle&=Ne^{+i\ve t}|p_z\rangle 
\begin{pmatrix}
p_z|n-1,\ell\rangle\\
i\sqrt{n}\vb |n,\ell\rangle\\
(\ve +m)|n-1,\ell\rangle\\
0
\end{pmatrix}
,\\
|\psi_4(n,\ell,p_z;t)\rangle&=Ne^{+i\ve t}|p_z\rangle 
\begin{pmatrix}
-i\sqrt{n}\vb |n-1,\ell\rangle\\
-p_z|n-1,\ell\rangle\\
0\\
(\ve +m)|n,\ell\rangle
\end{pmatrix}
. \label{sol4}
\end{align}
Except the ground state, each energy state is doubly degenerate with respect to the spin states of 
second order equation.  Each energy level is also infinitely degenerate with respect to integers $\ell$. 
The normalization constant $N$ is determined by 
\be 
\langle\psi_a(n,\ell,p_z)|\psi_a(n,\ell,p_z)\rangle\\ 
=2N^2\ve (\ve+m)=1 \ ({\rm for\ each}\ a). 
\ee
Hence, 
\be
N=[2\ve (\ve+m)]^{-1/2}.
\ee
This solution is a complete orthonormal, 
\be
\langle\psi_a(n,\ell,p_z)|\psi_a(n',\ell',p'_z)\rangle=\delta_{a,a'}\delta_{n,n'}\delta_{\ell,\ell'}\delta(p_z-p'_z),
\ee
and the completeness relation holds :
\be
\sum_{a,n,\ell}\int\! dp_z\; |\psi_a(n,\ell;t,p_z)\rangle\langle\psi_a^{\dagger}(n,\ell;t,p_z)|=\hat{I}, 
\ee
by use of relations (\ref{complete2}) and $\int\! dp_z|p_z\rangle\langle p_z|=\hat{I}$. 

Note that explicit representations for the $x,y$ variables have not been obtained so far.  
The explicit $xy$-coordinate representations are constructed by a projection of $\langle x,y|$ onto the above solution. 
Equally, an $\xi\bar{\xi}$-representation can be obtained by a projection of $\langle \xi,\bar{\xi}|$. 
In the latter case, the ground state $f(\xi,\bar{\xi})=\langle \xi,\bar{\xi}|0\rangle$ is easily found from two simple differential equations : 
\be
(\pd+\bar{\xi})f(\xi,\bar{\xi})=0,\quad (\bar{\pd}+\xi)f(\xi,\bar{\xi})=0.
\ee
The solution is 
\be
f(\xi,\bar{\xi})=\sqrt{\frac{2}{\pi}}\exp(-\xi\bar{\xi}). 
\ee
Here the normalization constant is determined by 
\be
\int\! d\mu(\xi)\;\overline{f(\xi,\bar{\xi})}f(\xi,\bar{\xi})=1,
\ee
where $d\mu(\xi)=d{\rm Re}\xi\,d{\rm Im}\xi$ is a measure for the integral. 
The normalization constant corresponding to the $xy$-representation is obtained 
by an additional factor $\vb/(2\sqrt2)$, i.e., 
\be
f(x,y)=\frac{\vb}{2\sqrt{\pi}}\exp[-\frac{\vb^2}{8}(x^2+y^2)].
\ee
The corresponding normalization is 
\be
\int\! dxdy\;\overline{f(x,y)}f(x,y)=1.
\ee
Using the creation operator $\beta^{\dagger}$ and the relation 
\be
e^{+\xi\bar{\xi}}\beta^{\dagger}e^{-\xi\bar{\xi}}
=e^{+\xi\bar{\xi}}\left(\frac{-\pd+\bar{\xi}}{\sqrt2}\right)e^{-\xi\bar{\xi}}=\frac{-\pd+2\bar{\xi}}{\sqrt2}, 
\ee
it is easy to show that the $\xi\bar{\xi}$-representation of the lowest eneregy state 
(Lowest Landau Level) $f^{LLL}_{\ell}(\xi,\bar{\xi})=\langle \xi,\bar{\xi}|n=0,\ell\rangle$ is 
\be
f^{LLL}_{\ell}(\xi,\bar{\xi})=\frac{(\sqrt2)^{\ell+1}}{\sqrt{\pi\ell!}}(\bar{\xi})^{\ell}e^{-\xi\bar{\xi}}.
\ee
The physical meaning of the quantum number $\ell$ is seen by evaluating 
the expectation value of the square of the distance from the origin 
$r^2=x^2+y^2=8\xi\bar{\xi}/\vb^2$ with respect to the lowest Landau levels as  
\bal
\langle\xi\bar{\xi} \rangle_{\ell}&= \int\! d\mu(\xi) \overline{f^{LLL}_{\ell}(\xi,\bar{\xi})}(\xi\bar{\xi})f^{LLL}_{\ell}(\xi,\bar{\xi})\\
&= \frac{2^{\ell+1}}{\pi \ell !} \int\! d\mu(\xi)\;(\xi\bar{\xi})^{\ell+1}e^{-2\xi\bar{\xi}}\\
&=\frac{2^{\ell+2}}{\ell !} \int^{\infty}_{0} \!dr\;r^{2\ell+3}e^{-2r^2}\\
&=\frac{\ell+1}{2}.
\end{align}
Here the spherical coordinates $\xi=r\exp(i\phi)$ is used to evaluate the integral. 
Thus the quantum number $\ell$ corresponds to discrete values of radius for 
the classical circular motion in a constant magnetic field. 

The complete orthonormal basis in the $\xi\bar{\xi}$-representation 
$\langle \xi,\bar{\xi}|n,\ell\rangle$ can be constructed in 
a similar manner, namely by applying $(\alpha^{\dagger})^n/\sqrt{n!}$ to $f^{LLL}_{\ell}(\xi,\bar{\xi})$. 
It is expressed in terms of hypergeometric functions. 
Such an explicit representation of the solution will not be used in the following.

\section{Second Quantization of the Dirac Field} 
The second quantization for the Dirac field in terms of the complete orthonormal basis is 
achieved as follows \cite{landauqed}. 
\be
\hat{\psi}(x)=\sum_{A}(b_A\psi^{+}_A(\v x)e^{-i\ve^+ t}+d^{\dagger}_A\psi^{-}_A(\v x)e^{+i\ve^-t} ), 
\ee
where $x$ represents $x^{\mu}=(t,\v x)$, and $A$ represents all collective indices, e.g. the energy level $n$, the spin $s$, and so on. 
$\psi^{+}_A(x)$ and $\psi^{-}_A(x)$ correspond to the positive and negative energy 
($\pm\ve^{\pm}$) solution of the single particle Dirac equation in external fields, i.e., 
\be
[\gamma^0(\pm\ve^{\pm})+\gamma^ip_i-e\slashed{A}-m]\psi_A^{\pm}(\v x)=0.
\ee
Similarly, the conjugate field $\bar{\psi}(x)=\psi^{\dagger}\gamma^0$ is quantized as 
\be
\hat{\bar{\psi}}(x)=\sum_{A}(b^{\dagger}_A\bar{\psi}^{+}_A(\v x)e^{+i\ve^+ t}
+d_A\bar{\psi}^{-}_A(\v x)e^{-i\ve^- t} ). 
\ee
In the same manner as the field quantization is done in terms of the plane wave basis,  
the equal time anti-commutation relation for the Dirac field is employed to satisfy the positivity of the eigenvalues of the Hamiltonian :  
\be
\left.\{\hat{\psi}_{\rho}(x),\hat{\psi}^{\dagger}_{\lambda}(x') \}\right|_{x_0=x'_0}=\delta_{\rho\lambda}\delta(\v x-\v x'), 
\ee
where indices $\rho$ and $\lambda$ denote components of the Dirac spinor. 
It then follows that the annihilation and creation operators $b_A$ and $b^{\dagger}_A$ for the positive energy 
particles satisfy 
\be
\{ b_A,b^{\dagger}_{A'}\}=\delta_{A,A'},  
\ee
and the annihilation and creation operators $d_A$ and $d^{\dagger}_A$ for the negative energy particles satisfy 
\be
\{ d_A,d^{\dagger}_{A'}\}=\delta_{A,A'}.  
\ee
In order to manifest diagram technics, define the propagator for the Dirac field by 
the vacuum expectation value of chronologically ordered field operators :
\be \label{propagator1}
iS(x;x')_{\rho\lambda}=\langle 0| T \hat{\psi}_{\rho}(x)\hat{\bar{\psi}}_{\lambda}(x')|0\rangle ,
\ee
where $T$ is the time ordering operation. For fermion field operators $\hat{A}$ and $\hat{B}$, 
the time ordering operation is defined as  
\be
T\v (\hat{A}(t)\hat{B}(t')\v ) \equiv 
\begin{cases}
+\hat{A}(t)\hat{B}(t')\ (t>t')\\
-\hat{B}(t')\hat{A}(t)\ (t<t')
\end{cases}
.
\ee
This definition can be extend to an arbitrary number of fermion field operators. In this case a sign factor 
$(-1)^n$ is assigned with $n$ corresponding to the number of permutations of operators in order to bring the original expression 
to the chronologically ordered expression.  It is also convenient to express the time ordering operation as 
\be
T\v (\hat{A}(t)\hat{B}(t')\v )=\Theta(t-t')\hat{A}(t)\hat{B}(t')-\Theta(t'-t)\hat{B}(t')\hat{A}(t),
\ee
where $\Theta(x)$ is the step function.  In the above definition of the propagator (\ref{propagator1}), 
$|0\rangle$ is the vacuum of the Dirac field, i.e.,
\be
b_A|0\rangle=d_A|0\rangle=0.
\ee
This vacuum is not to be confused with the vacuum used in the previous section to construct the solution of the Dirac equation. 
The quantized Dirac field $\hat{\psi}(x)$ satisfies the Dirac equation : 
\be \label{dfield}
(i\slashed{D}-m)\hat{\psi}(x)=0, 
\ee
where $D_{mu}=\pd_{\mu}+ieA_{\mu}$. 
The bare propagator (\ref{propagator1}) satisfies 
\be \label{bare1}
(i\slashed{D}-m)S(x;x')=\delta(x-x').  
\ee
It should be noted that the propagator depends on two points $x$ and $x'$ in general (not the difference $x-x'$), 
and hence the Fourier transform of the propagator does not take a simple form. 

\subsection{Propagator in a constant magnetic field}
Following the above argument, the propagator for electrons in a uniform magnetic field is constructed by 
\be
\hat{\psi}(x)=\langle x,y,z| \sum_{\ell,n=0}^{\infty}\sum_{s=1,2}\int\! dp_z\; 
\left( b_{\ell,n,s}(p_z)|\psi_s(\ell,n,p_z;t)\rangle+d^{\dagger}_{\ell,n,s}(p_z)|\psi_{s+2}(\ell,n,p_z;t)\rangle \right) .
\ee
Here the projection $\langle x,y,z|$ is taken in order for the field $\hat{\psi}(x)$ to be defined at all space-time points $x$. 
As was mentioned earlier, the $n=0$ case is allowed only for $s=2$. Using explicit expressions 
for the stationary solutions $|\psi_a\rangle$ (\ref{sol1}--\ref{sol4}), the propagator is 
\begin{multline}
iS(x;x')=\langle x,y| \sum_{\ell,n}\int\! \frac{dp_z}{2\pi}\; \frac{e^{ip_z \zeta}}{2\ve} 
\left[ \Theta(\tau)e^{-i\ve \tau}
\begin{pmatrix}
(\ve+m)D_n(\ell)&C_n(\ell,p_z)\\
-C_n(\ell,p_z)&-(\ve-m)D_n(\ell)
\end{pmatrix}
\right.
\\
-\Theta(-\tau)e^{+i\ve\tau}
\left.
\begin{pmatrix}
(\ve-m)D_n(\ell)&-C_n(\ell,p_z)\\
C_n(\ell,p_z)&-(\ve+m)D_n(\ell)
\end{pmatrix}
 \right]
|x',y'\rangle
,
\end{multline}
where $\zeta=z-z'$ and $\tau=t-t'$. To obtain the above expression, we also used 
the formula $\langle z|p_z\rangle=e^{izp_z}/\sqrt{2\pi}$ and its conjugate $\langle p_z|z\rangle=e^{-izp_z}/\sqrt{2\pi}$
The $2\times2$ matrices $C_n(\ell,p_z)$ and $D_n(\ell,p_z)$ are defined by 
\be
C_n(\ell,p_z)=
\begin{pmatrix}
-p_z\Lambda_{n-1,n-1}(\ell)&i\sqrt n \vb\Lambda_{n-1,n}(\ell)\\
-i\sqrt n \vb\Lambda_{n,n-1}(\ell)&p_z\Lambda_{n,n}(\ell)
\end{pmatrix}
, 
\ee
\be
D_n(\ell)=
\begin{pmatrix}
\Lambda_{n-1,n-1}(\ell)&0\\
0&\Lambda_{n,n}(\ell)
\end{pmatrix}
, 
\ee
together with a projection operator $\Lambda_{n,n'}(\ell)=|n,\ell\rangle\langle n',\ell |$. 
The step function $\Theta(x)$ has the Fourier transform of the form : 
\be
\Theta(x)=\frac{1}{2\pi i}\int_{-\infty}^{\infty}\!dk\;\frac{e^{ikx}}{k-i\epsilon},
\ee
where $\epsilon$ is a small positive number which goes to zero at the end of calculations. 
Using this formula, the propagator is expressed by  
\be \label{barepropagator}
S(x;x')=\langle x,y| \sum_{\ell,n}\int\! \frac{dp_0}{2\pi}\int\! \frac{dp_z}{2\pi}\; \frac{e^{-ip_0\tau+ip_z \zeta}}{p_0^2-\ve^2+i\epsilon} 
\begin{pmatrix}
(p_0+m)D_n&C_n\\
-C_n&-(p_0-m)D_n
\end{pmatrix}
|x',y'\rangle.
\ee
Using the relation $\langle z|p_z\rangle=e^{izp_z}/\sqrt{2\pi}$ and its conjugate $\langle p_z|z\rangle=e^{-izp_z}/\sqrt{2\pi}$, 
it is also possible to write the propagator as an operator in the form of $S$.  Then, the usual propagator 
is considered as a matrix element in space-time coordinates, i.e. $S(x;x')=\langle x|S|x'\rangle$. 
The operator $S$ is referred to by a Green's operator to distinguish it from propagators in particular 
representations such as the space-time coordinate, or the four momentum representations. 
The Green's operator satisfies 
\be
(\slashed{\Pi}-m)S=I, 
\ee
which is equivalent to eq.~(\ref{bare1}).
This description is more convenient in the following discussions. 

It is also convenient to obtain a coordinate representation for the electron propagator, which will be used later. 
The main concern is to evaluate 
\be
\langle x,y|\Lambda_{n,n'}(\ell)|x',y'\rangle, 
\ee
equivalently, 
\be
\langle \xi,\bar{\xi}|\Lambda_{n,n'}(\ell)|\xi',\bar{\xi'}\rangle. 
\ee
For this purpose it is easier to carry out the summation over $\ell$ first, then to evaluate the following function.  
\be
{\cal L}_{n,n'}(\xi,\bar{\xi};\xi'\bar{\xi'})\equiv \sum_{\ell}\langle \xi,\bar{\xi}|\Lambda_{n,n'}(\ell)|\xi',\bar{\xi'}\rangle. 
\ee
Since  
\be
\langle \xi,\bar{\xi}|n,\ell\rangle=\frac{(\sqrt2)^{\ell+1}}{\sqrt{\pi n!\ell!}}(\frac{-\bar{\pd}+\xi}{\sqrt2})^n (\bar{\xi})^{\ell}e^{-\xi\bar{\xi}}, 
\ee
the summation over $\ell$ gives   
\bal
{\cal L}_{n,n'}(\xi,\bar{\xi};\xi'\bar{\xi'})
&=\frac{2}{\pi\sqrt{n!n'!}}(\frac{-\bar{\pd}+\xi}{\sqrt2})^n(\frac{-\pd'+\bar{\xi'}}{\sqrt2})^{n'} 
e^{-\xi\bar{\xi}-\xi'\bar{\xi'}+2\bar{\xi}\xi'}\\
&= \frac{2e^{-\xi\bar{\xi}-\xi'\bar{\xi'}}}{\pi\sqrt{2^{n+n'}n!n'!}}(-\bar{\pd}+2\xi)^n(-\pd'+2\bar{\xi'})^{n'}e^{2\bar{\xi}\xi'}.
\end{align}
There are three cases for the values of $n$ and $n'$ : i) $n=n'$, ii) $n-1=n'$, and iii) $n+1=n'$. \\
We have for the case i),  
\bal
(-\bar{\pd}+2\xi)^n(-\pd'+2\bar{\xi'})^{n} e^{2\bar{\xi}\xi'}
&= (-\bar{\pd}+2\xi)^ne^{2\bar{\xi}\xi'}(-\pd'+2\bar{\xi'}-2\bar{\xi})^{n} \\
&=e^{2\bar{\xi}\xi'} (-\bar{\pd}+2\xi-2\xi')^n(2\bar{\xi'}-2\bar{\xi})^{n} \\
&=e^{2\bar{\xi}\xi'} \sum_{j=0}^{n}\binom{n}{j}(2\xi-2\xi')^{n-j}(-\bar{\pd})^j(2\bar{\xi'}-2\bar{\xi})^{n} \\
&=2^n n! e^{2\bar{\xi}\xi'} \sum_{j=0}^{n}\binom{n}{j}\frac{[-2(\xi-\xi')(\bar{\xi}-\bar{\xi'})]^j}{j!}.
\end{align}
Use the Laguerre polynomial defined by 
\be
L_n(x)= \sum_{j=0}^{n}\binom{n}{j}\frac{(-x)^j}{j!}, 
\ee
then 
\be
{\cal L}_{n,n}(\xi,\bar{\xi};\xi'\bar{\xi'})=\frac{2}{\pi}e^{-|\xi-\xi'|^2+\bar{\xi}\xi'-\xi\bar{\xi'}} L_n(2|\xi-\xi'|^2).
\ee
Similar calculation yields for the case ii) : 
\be \label{case2}
{\cal L}_{n+1,n}(\xi,\bar{\xi};\xi'\bar{\xi'})=\frac{2}{\pi}e^{-|\xi-\xi'|^2+\bar{\xi}\xi'-\xi\bar{\xi'}}\\
\frac{\sqrt2(\xi-\xi')}{\sqrt{n+1}} \left(L_n(2|\xi-\xi'|^2)+L^{(1)}_{n-1}(2|\xi-\xi'|^2)\right),
\ee
where $L^{(1)}_n(x)$ is the associate Laguerre polynomial defined by 
\be
L^{(1)}_n(x)= \sum_{j=0}^{n}\binom{n+1}{n-j}\frac{(-x)^j}{j!}. 
\ee
The case iii) is obtained by taking the complex conjugate of eq.~(\ref{case2}) together with 
exchanging $\xi\leftrightarrow\xi'$ and $\bar{\xi}\leftrightarrow\bar{\xi'}$.
In the same way as before, the following convention is understood,  
\be
L_{-n}(x)=L^{(1)}_{-n}(x)=0\quad {\rm for}\ n<0.
\ee
Also an additional factor in order to get the $xy$-representation should be remembered. 

Therefore, the explicit coordinates representation of the propagator is  
\begin{multline}
S(x;x')=\frac{2}{\pi}\sum_n\int\! \frac{dp_0}{2\pi}\int\! \frac{dp_z}{2\pi}\;
\frac{e^{-ip_0\tau+ip_z \zeta-|\xi-\xi'|^2+\bar{\xi}\xi'-\xi\bar{\xi'}}}{p_0^2-\ve^2+i\epsilon} \\
\times
\begin{pmatrix}
(p_0+m)\tilde{D}_n&\tilde{C}_n\\
-\tilde{C}_n&-(p_0-m)\tilde{D}_n
\end{pmatrix}
,
\end{multline}
with the definition of $2\times2$ matrices : 
\be
\tilde{C}_n=-
\begin{pmatrix}
p_z L_{n-1}&i\sqrt 2 \vb(\bar{\xi}-\bar{\xi'})(L_{n-1}+L^{(1)}_{n-2})\\
i\sqrt 2 \vb(\xi-\xi')(L_{n-1}+L^{(1)}_{n-2})&-p_z L_{n}
\end{pmatrix}
, 
\ee
\be
\tilde{D}_n=
\begin{pmatrix}
L_{n-1}&0\\
0&L_{n}
\end{pmatrix}
.
\ee
In this expression the argument of the Laguerre polynomials is a function of 
$2|\xi-\xi'|^2$, i.e. $L_n^{(0,1)}(2|\xi-\xi'|^2)$. Therefore the only term, which 
depends on two points $\xi$ and $\xi'$, appears as an exponential factor $\exp(\bar{\xi}\xi'-\xi\bar{\xi'})$.
This, of course, agrees with the general statement that the electron propagator is gauge dependent object 
with an additional phase factor of the form : $\exp\v (i e \int^{x}_{x'} dx^{\mu} A_{\mu} (x^{\mu}) \v )$. 

\subsection{Photon propagator}

The propagator for a quantized electromagnetic field, the photon propagator, is defined by 
\be
-i D_{\mu\nu}(x-x')=\langle 0| T\hat{A}_{\mu}(x)\hat{A}_{\nu}(x')|0\rangle, 
\ee
where $\hat{A}_{\mu}(x)$ is the second quantized field for photons. 
Associated with the gauge degree of freedom of $\hat{A}_{\mu}(x)$, 
the photon propagator is not determined uniquely. 
The most common gauge is the Feynman gauge, which is expressed in Heaviside-Lorentz units as, 
\be
D_{\mu\nu}(x-x')=-\frac{1}{4\pi^2}\frac{g_{\mu\nu}}{(x-x')^2-i\epsilon}.  
\ee
Its Fourier transform is 
\be
D_{\mu\nu}(k)=\frac{g_{\mu\nu}}{k_0^2-\v k^2+i\epsilon}.
\ee
In the following, the Coulomb gauge is also employed. 
The Fourier transform of the photon propagator in the Coulomb gauge is 
\bal
D_{00}(k)&=-\frac{1}{\v k^2},\\
D_{0i}(k)&=0,\\ 
D_{ij}(k)&=-\frac{1}{k_0^2-\v k^2+i\epsilon}(\delta_{ij}-\frac{k_ik_j}{\v k^2}).
\end{align}

\section{Modified Dirac Equation} 
The mass operator $\cal M$ is defined by a summation of all irreducible diagrams which 
contain two external electron lines. Irreducible diagrams are defined by 
diagrams which cannot be reduced to a sum of simpler diagrams by cutting lines. 
To the lowest order correction, it is 
represented by this diagram 
\begin{fmffile}{second}
\begin{equation}
\label{selfenergy}
\parbox{30mm}{
\begin{fmfgraph}(60,40)
	\fmfleft{i}
	\fmfright{o}
	\fmf{plain}{i,v1}
	\fmf{fermion,tension=1/3}{v1,v2}
	\fmf{photon,left,tension=1/3}{v1,v2}
	\fmf{plain}{v2,o}
	\fmfdot{v1,v2}
\end{fmfgraph}
}
.
\end{equation}
\end{fmffile}
Let $S$ be the bare Green's operator for electrons, 
which is constructed directly from the solution of the single particle Dirac equation in external fields. 
Let $\cal G$ be the exact Green's operator for electrons, 
then it satisfies the following relation among the bare Green's operator and the mass operator : 
\begin{fmffile}{sdeq}
\be
\parbox{20mm}{
\begin{fmfgraph}(50,35)
	\fmfleft{i}
	\fmfright{o}
	\fmf{heavy}{i,o}
\end{fmfgraph}
}
=
\parbox{20mm}{
\begin{fmfgraph}(50,35)
	\fmfleft{i}
	\fmfright{o}
	\fmf{fermion}{i,o}
\end{fmfgraph}
}
\\+
\parbox{20mm}{
\begin{fmfgraph}(50,35)
	\fmfleft{i}
	\fmfright{o}
	\fmf{fermion}{i,v}
	\fmf{fermion}{v,o}
	\fmfblob{0.2w}{v}
\end{fmfgraph}
}
+
\parbox{20mm}{
\begin{fmfgraph}(50,35)
	\fmfleft{i}
	\fmfright{o}
	\fmf{fermion}{i,v1}
	\fmf{fermion,tension=2/3}{v1,v2}
	\fmfblob{0.15w}{v1}
	\fmf{fermion}{v2,o}
	\fmfblob{0.15w}{v2}
\end{fmfgraph}}
+\cdots .
\ee
\end{fmffile}
In the above diagrams, the thick line represents the exact Green's operator $i\cal G$, 
and each shadowed circle corresponds to the mass operator $-i\cal M$. 
It should be emphasized that this is obtained under an assumption that the exact Green's operator 
can be decomposed into a summation of irreducible diagrams. 
Express the above diagram relation as an equation by   
\bal
{\cal G}&=S+S{\cal M}S+S{\cal M}S{\cal M}S+\cdots\\
&=S+S{\cal M}(S+S{\cal M}S+\cdots)\\
&=S+S{\cal MG},
\end{align}
thus 
\be \label{SJ1}
{\cal G}-S=S{\cal MG}. 
\ee
Inserting the following identity, this can be expressed in the coordinate representation  
\be
\int\! dx\;|x\rangle\langle x|=I, 
\ee
\be
{\cal G}(x;x')-S(x;x')\\=\int\! dx_1\int\! dx_2\;S(x;x_1){\cal M}(x_1;x_2){\cal G}(x_2;x').
\ee
Useful relations are derived by applying an operator $\slashed{\Pi}=\slashed{P}-e\slashed{A}$ to eq.~(\ref{SJ1}) from the left :  
\bal
\slashed{\Pi}{\cal G}&=\slashed{\Pi}S+\slashed{\Pi}S{\cal MG}\\
&=(m+I)S+(m+I)S{\cal MG}\\
&=m{\cal M}+I+{\cal MG}, 
\end{align}
where $(\slashed{\Pi}-m)S=I$ is used. 
Define a new Green's operator $\tilde{\cal M}$ by 
\be
\tilde{\cal M}=m+{\cal M}, 
\ee
where the coordinate representation is 
\be
\tilde{\cal M}(x;x')=m\delta(x-x')+{\cal M}(x;x').
\ee
The exact Green's operator satisfies 
\be \label{SJ2}
(\slashed{\Pi}-\tilde{\cal M}){\cal G}=I.
\ee
In the coordinate representation, 
\be \label{SJ3}
i\slashed{D}{\cal G}(x;x')-\int\! dx_1\;\tilde{\cal M}(x;x_1){\cal G}(x_1;x')=\delta(x-x').
\ee
Therefore it is found that the exact Green's operator satisfies the modified Dirac equation (\ref{SJ2}) or (\ref{SJ3}), 
where the mass $m$ is replaced with the operator $\tilde{{\cal M}}=m+{\cal M}$ \cite{schwinger51}. 
Note that the poles of the propagator correspond to the energy levels of the system as seen by 
the expression (\ref{barepropagator}). 
Then this modified Dirac equation for the exact Green's operator is particularly important, 
since the poles of the exact propagator contains information about the exact energy 
levels of the system.  Inverting the argument to derive eq.~(\ref{bare1}), the exact propagator 
assures the exact solution $|\Psi\rangle$ of the modified Dirac equation :  
\be \label{exact1}
(\slashed{\Pi}-\tilde{\cal M})|\Psi\rangle=0,  
\ee
or, 
\be
i\slashed{D}\Psi(x)-\int\! dx'\;\tilde{\cal M}(x;x')\Psi(x')=0.
\ee
Choose the positive energy $\even$ solution $|\Psi_{\even}\rangle$, i.e.,
\be
P_0 |\Psi_{\even}\rangle=\even|\Psi_{\even}\rangle,
\ee
then eq.~(\ref{exact1}) is rewritten as 
\be \label{exact2}
(\gamma^0\even+\gamma^iP_i-e\slashed{A}-\tilde{\cal M})|\Psi_{\even}\rangle=0.
\ee
Assume that the exact eigenstate is slightly different from the eigenstate of the original Dirac equation 
due to radiative corrections. Under this assumption the eigenstae is  written as 
\be \label{pert1}
|\Psi_{\even}\rangle=|\psi_{E}\rangle+|\delta\psi\rangle, 
\ee
where $|\psi_{E}\rangle$ satisfies 
\be \label{unpert1}
(\gamma^0E+\gamma^iP_i-e\slashed{A}-m)|\psi_{E}\rangle=0.
\ee
By our assumption, $\Delta E=\even-E$ and $| |\delta\psi\rangle |$ are small. 
Substituting (\ref{pert1}) into eq.~(\ref{exact2}) and using (\ref{unpert1}),  
\be \label{energyshift}
(\even-E)|\psi_{E}\rangle=
\gamma^0{\cal M}|\psi_{E}\rangle \\
-\gamma^0[(\gamma^0\even+\gamma^iP_i-e\slashed{A}-\tilde{\cal M}] |\delta\psi\rangle.
\ee
Multiply both sides by$\langle \psi_{E}|$ from the left, 
\be
\even-E=\langle \psi_{E}|\gamma^0{\cal M}|\psi_{E}\rangle \\
-\langle \psi_{E}|\gamma^0[(\gamma^0\even+\gamma^iP_i-e\slashed{A}-\tilde{\cal M}] |\delta\psi\rangle.
\ee
By egarding the operators in the last term as acting on $\langle \psi_{E}|$ and using the conjugate Dirac equation, 
\be
\langle \psi_{E}|\gamma^0 [\gamma^0(-E)+\gamma^iP_i+e\slashed{A}+m]=0,
\ee
eq.~(\ref{energyshift}) becomes 
\be
\even-E=\langle \psi_{E}|\gamma^0{\cal M}|\psi_{E}\rangle \\
+\langle \psi_{E}|\gamma^0[-\gamma^0(\even+E)+2m+2e\slashed{A}+{\cal M}] |\delta\psi\rangle.
\ee
The second term in the right hand side is regarded as a next order contribution.  
Therefore the energy shift is found in the perturbation expansion to the lowest order as 
\be
\even-E=\langle \psi_{E}|\gamma^0{\cal M}|\psi_{E}\rangle.
\ee
In the coordinate representation, this is written 
\be
\even-E=\int\! d^3x \int\! d^3x'\;\bar{\psi}_{E}(\v x) {\cal M}(E |\v x;\v x')\psi_{E}(\v x'), 
\ee
where ${\cal M}(E|\v x;\v x')$ is the Fourier transform of the mass operator in the coordinate representation 
with respect to time variable, i.e., 
\be
{\cal M}(E|\v x;\v x')=\left.\int\! d\tau\;e^{i\omega\tau} {\cal M}(x;x')\ \right|_{\omega=E}.
\ee
A time dependence of the form $\tau=t-t'$ of the mass operator is guaranteed for homogeneous external fields case. 

\section{One Loop Correction to the Mass Operator I} 
The one loop correction to the mass operator corresponding to the diagram (\ref{selfenergy}) is calculated in this section. 
To this end it is useful to adopt the mixed representation for the propagators, 
namely momentum space for the time and $z$-component of coordinates, while we use the position coordinates for the $x,y$-variables. 
The reason for this choice is clear, since the bare electron propagator 
depends on $t-t'$ and $z-z'$. Therefore, ordinary Feyman rules in momentum space is applied for the 
$p_0$ and $p_z$ variables. In order to review notations, a summary for the propagators in this representation is given below. 
The bare electron propagator is 
\be
S(p_0,p_z|\xi,\bar{\xi};\xi',\bar{\xi'})\\=\frac{2}{\pi}\sum_n
\frac{e^{-|\xi-\xi'|^2+\bar{\xi}\xi'-\xi\bar{\xi'}}}{p_0^2-\ve ^2+i\epsilon}\Gamma_n(p_0,p_z|\xi-\xi',\bar{\xi}-\bar{\xi'}), 
\ee
where the energy levels are $\ve =\sqrt{m^2+p_z^2+\vb^2n}$, and a $4\times4$ matrix $\Gamma_n$ is 
\be
\Gamma_n(p_0,p_z|\xi-\xi',\bar{\xi}-\bar{\xi'})=
\begin{pmatrix}
(p_0+m)\tilde{D}_n&\tilde{C}_n\\
-\tilde{C}_n&-(p_0-m)\tilde{D}_n
\end{pmatrix}
,
\ee
\be
\tilde{C}_n(p_z|\xi,\bar{\xi};\xi',\bar{\xi'})=-
\begin{pmatrix}
p_z L_{n-1}&i\sqrt 2 \vb(\bar{\xi}-\bar{\xi'})(L_{n-1}+L^{(1)}_{n-2})\\
i\sqrt 2 \vb(\xi-\xi')(L_{n-1}+L^{(1)}_{n-2})&-p_z L_{n}
\end{pmatrix}
, 
\ee
\be
\tilde{D}_n(|\xi-\xi'|^2)=
\begin{pmatrix}
L_{n-1}(2|\xi-\xi'|^2)&0\\
0&L_{n}(2|\xi-\xi'|^2)
\end{pmatrix}
.
\ee
The photon propagator in Coulomb gauge is 
\be
D_{00}(k_0,k_z|\xi-\xi',\bar{\xi}-\bar{\xi'})=-\int\! \frac{d^2k_{\bot}}{(2\pi)^2}\; 
\frac{e^{i\bar{\kappa}(\xi-\xi')+i\kappa(\bar{\xi}-\bar{\xi'})}}{ k^2},
\ee
\be
D_{0i}(k_0,k_z|\xi-\xi',\bar{\xi}-\bar{\xi'})=0,
\ee
\be
D_{ij}(k_0,k_z|\xi-\xi',\bar{\xi}-\bar{\xi'})\\=-\int\! \frac{d^2k_{\bot}}{(2\pi)^2}\; 
\frac{e^{i\bar{\kappa}(\xi-\xi')+i\kappa(\bar{\xi}-\bar{\xi'})}}{k_0^2- k^2+i\epsilon}(\delta_{ij}-\frac{k_ik_j}{ k^2}),
\ee
where $d^2k_{\bot}=dk_xdk_y$, $\kappa=\sqrt2(k_x+ik_y)/\vb$, and $k=|\v k|$. 

In the first approximation, the mass operator is approximated by the diagram (\ref{selfenergy}), 
\begin{multline}
-i{\cal M}(p_0,p_z|\xi-\xi',\bar{\xi}-\bar{\xi'})\\=-e^2\int\! \frac{dk_0}{2\pi}\int\! \frac{dk_z}{2\pi}\; 
\gamma^{\mu} D_{\mu\nu}(k_0,k_z|\xi-\xi',\bar{\xi}-\bar{\xi'})S(p_0-k_0,p_z-k_z|\xi,\bar{\xi};\xi',\bar{\xi'})\gamma^{\nu}.
\end{multline}
As was shown in the previous section, the energy shift of the ground state due to the radiative correction is obtained by 
the expectation value of the mass operator with respect to the ground state. 
The ground state ($E=m,p_z=0$) of the electron in a constant magnetic field in this mixed representation is given by 
\be
\psi_g(|\xi|^2)=\langle\xi,\bar{\xi}|\psi_g\rangle=\sqrt{\frac{2}{\pi}}e^{-|\xi|^2}w_2|p_z=0\rangle, 
\ee
where $w_2^{\dagger}=(0,1,0,0)$ is the eigenspinor, and a phase factor $\exp(-imt)$ is dropped. 
To reduce the amount of calculations, we also choose $\ell=0$ state, 
but all other choices of the lowest Landau level give the same result.  
This comes from the fact that the ground state degeneracy is not affected by 
the first order radiative correction. 
Therefore, the energy shift $\Delta E_g$ for the ground state is 
\bal
\Delta E_g&=\left.\langle \psi_g|\gamma^0 {\cal M}|\psi_g\rangle \right|_{p_0=m,p_z=0}\\ 
&=\int\! d\mu(\xi)\int\! d\mu(\xi')\;\psi_g(|\xi|^2)\psi_g(|\xi'|^2)\langle \xi,\bar{\xi}| \gamma^0 {\cal M}|\xi',\bar{\xi'}\rangle|_{E=m,p_z=0}\\ \nonumber
&=\frac{4ie^2}{\pi^2}\sum_n \int\! d\mu(\xi)\int\! d\mu(\xi')\;e^{-|\xi|^2-|\xi'|^2}\int\! \frac{d^4k}{(2\pi)^4}\;
e^{i\bar{\kappa}(\xi-\xi')+i\kappa(\bar{\xi}-\bar{\xi'})}\\ \nonumber
&\hspace{0.5cm}\times \frac{e^{-|\xi-\xi'|^2+\bar{\xi}\xi'-\xi\bar{\xi'}}}{(m-k_0)^2-\ve^2+i\epsilon} 
\left(\frac{1}{k^2}\bar{w_2}\gamma^0\Gamma_n(m-k_0,-k_z|\xi-\xi',\bar{\xi}-\bar{\xi'})\gamma^0w_2\right. \\ 
&\hspace{1cm}+\left.\frac{\delta_{ij}-k_ik_j/ k^2}{k_0^2- k^2+i\epsilon} 
\bar{w_2}\gamma^i\Gamma_n(m-k_0,-k_z|\xi-\xi',\bar{\xi}-\bar{\xi'})\gamma^jw_2\right),
\end{align}
where $\ve=\sqrt{m^2+(-k_z)^2+\vb^2n}$. 
Changing integration variables to $\xi$ and $\chi=\xi-\xi'$, 
\begin{multline}
\Delta E_g=
\frac{4ie^2}{\pi^2}\sum_n \int\! d\mu(\xi)\int\! d\mu(\chi)\int\! \frac{d^4k}{(2\pi)^4}\;
\frac{e^{-2|\xi|^2+2\bar{\chi}\xi-2|\chi|^2+i\bar{\kappa}\chi+i\kappa\bar{\chi}}}{(m-k_0)^2-\ve^2+i\epsilon}\\
\times\left(\frac{1}{ k^2}\bar{w_2}\gamma^0\Gamma_n(m-k_0,-k_z|\chi,\bar{\chi})\gamma^0w_2
+\frac{\delta_{ij}-k_ik_j/ k^2}{k_0^2- k^2+i\epsilon} 
\bar{w_2}\gamma^i\Gamma_n(m-k_0,-k_z|\chi,\bar{\chi})\gamma^jw_2\right).
\end{multline} 
Carry out the integration over $\xi$, which is an ordinary gaussian integral, to get 
\begin{multline}
\Delta E_g=
\frac{2ie^2}{\pi}\sum_n \int\! d\mu(\chi)\int\! \frac{d^4k}{(2\pi)^4}\;
\frac{e^{-2|\chi|^2+i\bar{\kappa}\chi+i\kappa\bar{\chi}}}{(m-k_0)^2-\ve^2+i\epsilon}\\
\times\left(\frac{1}{ k^2}\bar{w_2}\gamma^0\Gamma_n(m-k_0,-k_z|\chi,\bar{\chi})\gamma^0w_2 
+\frac{\delta_{ij}-k_ik_j/ k^2}{k_0^2- k^2+i\epsilon} 
\bar{w_2}\gamma^i\Gamma_n(m-k_0,-k_z|\chi,\bar{\chi})\gamma^jw_2\right).
\end{multline}
The matrix elements are evaluated as  
\be
\bar{w_2}\gamma^0\Gamma_n(m-k_0,-k_z|\chi,\bar{\chi})\gamma^0w_2
=-(k_0-2m)L_n(2|\chi|^2).
\ee
Using the formula,
\be
\sigma^i 
\begin{pmatrix}
a&0\\
0&d
\end{pmatrix}
\sigma^i=
\begin{pmatrix}
a+2d&0\\
0&d+2a
\end{pmatrix}
,
\ee
\be
\bar{w_2}\gamma^i\Gamma_n(m-k_0,-k_z|\chi,\bar{\chi})\gamma^iw_2\\
=-k_0\v (L_n(2|\chi|^2)+2L_{n-1}(2|\chi|^2)\v ).
\ee
Using another formula, 
\be
k_ik_j\sigma^i 
\begin{pmatrix}
a&0\\
0&d
\end{pmatrix}
\sigma^j= k^2
\begin{pmatrix}
d&0\\
0&a
\end{pmatrix}
+(a-d)k_3k_i\sigma^i,
\ee
yields 
\begin{multline}
\frac{k_ik_j}{ k^2}\bar{w_2}\gamma^i\Gamma_n(m-k_0,-k_z|\chi,\bar{\chi})\gamma^jw_2\\
=-\frac{k_0}{ k^2}[ k^2L_{n-1}(2|\chi|^2)+k_z^2\v (L_{n}(2|\chi|^2)-L_{n-1}(2|\chi|^2)\v )].
\end{multline}
Hence,
\begin{multline}
\Delta E_g=
\frac{-2ie^2}{\pi}\sum_n \int\! d\mu(\chi)\int\! \frac{d^4k}{(2\pi)^4}\;
\frac{e^{-2|\chi|^2+i\bar{\kappa}\chi+i\kappa\bar{\chi}}}{(m-k_0)^2-\ve^2+i\epsilon}\;\frac{1}{ k^2}\\
\times\left\{(k_0-2m)L_n+\frac{k_0}{k_0^2- k^2+i\epsilon}\left[ k^2(L_n+L_{n-1})-k_z^2(L_n-L_{n-1}) \right] \right\}.
\end{multline}
The integration for $\chi$ is done with the aid of the formula 
\be
\int_{0}^{\infty}\!dx\;xe^{-x^2}L_n(x^2)J_0(xy)=\frac {e^{-y^2/4}}{2} \frac{(y/2)^{2n}}{n!},
\ee
where $J_0(x)$ is the Bessel function of order zero, which has an integral representation of 
\be
J_0(x)=\frac{1}{2\pi}\int_{-\pi}^{\pi}\!d\phi\;e^{-ix\sin\phi}.
\ee
Going to spherical coordinates $\chi=r\exp(i\phi)$, the integral in question is evaluated as follows. 
\bal
I_n&\equiv\int\! d\mu(\chi)e^{-2|\chi|^2+i\bar{\kappa}\chi+i\kappa\bar{\chi}}L_n(2|\chi|^2)\\
&=2\pi \int_0^{\infty}\!dr\;re^{-2r^2}L_n(2r^2)J_0(2|\kappa|r)\\
&=\frac{\pi}{2}e^{-k_{\bot}^2/\vb^2}e_n(k_{\bot}^2/\vb^2), 
\end{align}
where $k_{\bot}=\sqrt{k_x^2+k_y^2}$ and the function $e_n(x)$ is defined by 
\be
e_n(x)=\frac{x^n}{n!}.
\ee
Then $\chi$ integration leads to  
\bml
\Delta E_g=
-ie^2\sum_n\int\! \frac{d^4k}{(2\pi)^4}\;\frac{e^{-k_{\bot}^2/\vb^2}}{(m-k_0)^2-\ve^2+i\epsilon}\;\frac{1}{ k^2}\\
\times \left\{(k_0-2m)e_n(k_{\bot}^2/\vb^2)+\frac{k_0\left[k_{\bot}^2e_n(k_{\bot}^2/\vb^2)
+(k^2+k_z^2)e_{n-1}(k_{\bot}^2/\vb^2) \right]}{k_0^2- k^2+i\epsilon} \right\}.
\end{multline}
Lastly, the contour integrals to evaluate $k_0$ integral are used,  
\bal
&\int\!\frac{dk_0}{2\pi i}\;\frac{k_0-2m}{(m-k_0)^2-\ve^2+i\epsilon}=\frac{m}{2\ve},\\ \nonumber
&\int\!\frac{dk_0}{2\pi i}\;\frac{k_0}{[(m-k_0)^2-\ve^2+i\epsilon][k_0^2- k^2+i\epsilon]}=\frac{m}{2\ve}\frac{1}{(\ve+k)^2-m^2}.
\end{align}
Therefore, the energy shift of the ground state is found in the lowest correction as  
\be \label{eshift}
\Delta E_g=
2\pi m\alpha\sum_{n=0}^{\infty}\int\! \frac{d^3k}{(2\pi)^3}\;\frac{e^{-k_{\bot}^2/\vb^2}}{\ve k^2}\\
\left[e_n\left(\frac{k_{\bot}^2}{\vb^2}\right)+\frac{k_{\bot}^2e_n\left(\frac{k_{\bot}^2}{\vb^2}\right)+(k^2+k_z^2)e_{n-1}\left(\frac{k_{\bot}^2}{\vb^2}\right)}{(\ve+k)^2-m^2}\right], 
\ee
where $\alpha=e^2/(4\pi\hbar c)$ is the fine structure constant in Heaviside-Lorentz units, 
and the same convention $e_{-1}(k_{\bot}^2/\vb^2)=0$ should be noted. 

This integral (\ref{eshift}) exactly agrees with the one obtained by Luttinger based on different arguments \cite{luttinger}. 
Note that this integral is essentially two dimensional since it does not depend on the direction of $\v{k_{\bot}}=(k_x,k_y)$. 
Because of its complexity, it is unlikely that we can obtain an analytic expression for this integral, 
especially the second term in the square bracket.

\section{One Loop Correction to the Mass Operator II} 
In this section an alternative expression is given for the first order radiative correction to 
the ground state energy. By reexamining Luttinger's result, it becomes clear that his 
expression is equivalent to Schwinger's expression. This will be shown briefly below. 

\subsection{Comparison to the proper time representation}
The photon propagator in the Feynman gauge is employed instead of the Coulomb gauge in this section. 
Since the calculation is almost same as was done for the Coulomb gauge, we only show the result. 
After integrating over holomorphic coordinate variables $\xi$ and $\xi'$, the energy shift for the ground state is  
\be \label{eshiftfey1}
\Delta E_g=
-2ie^2\sum_n\int\! \frac{d^4k}{(2\pi)^4}\;e^{-k_{\bot}^2/\vb^2}
\frac{me_n(k_{\bot}^2/\vb^2)+k_0e_{n-1}(k_{\bot}^2/\vb^2)}{[(m-k_0)^2-\ve^2+i\epsilon][k_0^2- k^2+i\epsilon]}.
\ee
Using Schwinger's trick to bring denominators into the exponent, i.e. $i/(a+i\epsilon)=\int_0^{\infty}d\alpha e^{i\alpha(a+i\epsilon)}$, 
eq.~(\ref{eshiftfey1}) becomes 
\begin{multline}
\Delta E_g=
(-i)^32e^2\sum_n\int\!\frac{d^4k}{(2\pi)^4}\int_0^{\infty}\!d\alpha_1\int_0^{\infty}\!d\alpha_2\;\left(me_n+k_0e_{n-1} \right)\\
\times \exp\{-k_{\bot}^2/\vb^2 +i\alpha_1[(m-k_0)^2-\ve^2]+i\alpha_2[k_0^2- k^2]\}.
\end{multline}
A small positive $\epsilon$ for the integral convergency should be reminded. 
A summation over Landau Levels $n=0,1,2,\cdots$ can be carried out at this point, 
which is just an exponential series;   
\bml
\Delta E_g=
(-i)^32e^2\int\! \frac{d^4k}{(2\pi)^4}\int\!d\alpha_1\int\!d\alpha_2\;(m+k_0e^{-i\alpha_1\vb^2}) \\ 
\times\exp\left[i(\alpha_1+\alpha_2)k_0^2-2i\alpha_1mk_0-(1-e^{i\alpha_1\vb^2}+i\alpha_2\vb^2)k_{\bot}^2/\vb^2-i(\alpha_1+\alpha_2)k_z^2 \right].
\end{multline}
Lastly, integrals over the four momentum is done to yield 
\be
\Delta E_g=
\frac{im\alpha}{2\pi}\int\!d\alpha_1\int\!d\alpha_2\;
\frac{\vb^2(\alpha_1+\alpha_2+\alpha_2e^{-i\alpha_1\vb^2})}{(\alpha_1+\alpha_2)^2(1-e^{i\alpha_1\vb^2}+i\alpha_2\vb^2)}
\exp(-i\frac{m^2\alpha_1^2}{\alpha_1+\alpha_2}) .
\ee
To see the exact correspondence, we change integration variables $\alpha_1,\alpha_2$ to 
$x=\vb^2\alpha_2/2$ and $u=\alpha_1/(\alpha_1+\alpha_2)$.  Thus, 
\be \label{eshiftsch}
\Delta E_g=
\frac{m\alpha}{2\pi}\int_0^{\infty}\!\frac{dx}{x}\int_0^1\!du\;
\frac{1+ue^{-2ix}}{1-u+ue^{-ix}\sin x/x}\exp(-\frac{i2m^2ux}{\vb^2}). 
\ee
This is identical to the expression obtained by Schwinger except for the contact term in his expression \cite{psf}. 
This contact term will be discussed in Sec.~8.  

It is also straightforward to check eq.~(\ref{eshiftfey1}) is indeed equivalent to eq.~(\ref{eshift}) by 
integrating over $k_0$ in eq.~(\ref{eshiftfey1}). This leads to 
\be \label{eshiftfey2}
\Delta E_g=
4\pi m\alpha\sum_{n=0}^{\infty}\int\! \frac{d^3k}{(2\pi)^3}\;\frac{e^{-k_{\bot}^2/\vb^2}}{(\ve+k)^2-m^2}\\
\times\left[(\frac{1}{\ve}+\frac1k)e_n(k_{\bot}^2/\vb^2)+\frac{1}{\ve}e_{n-1}(k_{\bot}^2/\vb^2)\right], 
\ee
which is identical to eq.~(\ref{eshift}) with the aid of relation $\ve^2=m^2+k_z^2+n\vb^2$. 

\subsection{Alternative representation}
We next derive an alternative representation for the energy shift $\Delta E_g$. 
The derivation is based on use of the following identity in eq.~(\ref{eshiftfey1}), 
\be
\frac{1}{AB}=\int_0^1\!dt \frac{1}{[t A+(1-t)B]^2}. 
\ee
After rewriting the product of two terms in the denominator of (\ref{eshiftfey1}) using this trick, 
elementary integrations over $k_0$ and $k_z$ yields 
\be \label{eshiftfey3}
\Delta E_g=
\frac{m\alpha}{4\pi^2} \sum_{n=0}^{\infty}\int\!dk^2_{\bot}\int_0^1\!dt\;
\frac{e_n(k_{\bot}^2/\vb^2)+t e_{n-1}(k_{\bot}^2/\vb^2)}{(1-t)k^2_{\bot}+m^2t^2+n\vb^2t}e^{-k^2_{\bot}/\vb^2}.
\ee
We can further proceed $t$ integral in a simple manner.  
Define 
\be
I(\omega;n\eta)=\frac 12\int_0^1\!dt\;\frac{1}{t^2-(\omega-n\eta)t+\omega}, 
\ee
with $\omega=k^2_{\bot}/m^2$ and $\eta=\vb^2/m^2$. 
Then, depending on the value of $\omega$ (or $k^2_{\bot}$), $t$ integral $I(\omega;n\eta)$ 
can be expressed in terms of inverse of 
either tangent or hyperbolic tangent as follows. 
\be
I(\omega;n\eta)=
\begin{cases}
\frac{1}{\sqrt{D_{\omega}}} \tanh^{-1} (\frac{\sqrt{D_{\omega}}}{\omega+n\eta})\quad; 0\le\omega \le\omega_-\ {\rm or}\ \omega_+\le\omega\\
\frac{1}{\sqrt{-D_{\omega}}} \tan^{-1} (\frac{\sqrt{-D_{\omega}}}{\omega+n\eta})\quad; \omega_- \le\omega\le\omega_+ .
\end{cases}
\ee
Here $D_{\omega}$ and $\omega_{\pm}$ are defined as 
\bal
D_{\omega}&=\omega^2-2(n\eta+2)\omega+n^2\eta^2=(\omega-\omega_+)(\omega-\omega_-), \\
\omega_{\pm}&=(\sqrt{n\eta+1}\pm1)^2.
\end{align}
Using this definition for $I^{(n)}(\omega;\eta)$, we arrive at the following alternative expression for the energy shift : 
\begin{multline} \label{eshiftj}
\Delta E_g=
\frac{m\alpha}{2\pi} \sum_{n=0}^{\infty}\int_0^{\infty}\!d\omega\;
e^{-\omega/\eta}\left\{\left[2 e_n(\frac{\omega}{\eta})+(\omega-n\eta) e_{n-1}(\frac{\omega}{\eta})\right]I(\omega;n\eta)\right.\\
+\frac 12\left. \log(\frac{n\eta+1}{\omega})e_{n-1}(\frac{\omega}{\eta}) \right\}.
\end{multline}
Therefore, we have obtained one integral over the transverse momentum $|k_{\bot}|$ and summation over 
the Landau levels $n$. This expression (\ref{eshiftj}) may have an advantage compared to the standard 
double integral expression (\ref{eshiftsch}) particularly for numerical purpose. 

\section{Anomalous Magnetic Moment} 
We would now like to bring our attention to discussion on the anomalous magnetic moment of an electron 
for the weak field case. 
The magnetic moment $\mu_e$ of the electron is defined by 
\be \label{defmu}
\mu_e=-\left.\frac{\pd E_g}{\pd B}\right|_{B=0}.
\ee
In other words, by applying a weak external magnetic field to the electron, the ground state energy is shifted 
by amount $-\mu_e B$. Therefore the main concern is to examine the above energy shift in the weak magnetic field limit, 
or a small $\vb=\sqrt{-2eB}$ limit.  Our derivation of the anomalous magnetic moment is based on 
the method used by Luttinger \cite{luttinger}. To appreciate his elegant derivation, 
we use the expression (\ref{eshift}) obtained first by him. 
Eq.~(\ref{eshift}) is rewritten as 
\begin{multline} \label{eshift2}
\Delta E_g=\frac{m\alpha}{2\pi}\sum_{n=0}^{\infty}\int_{0}^{\infty}\!dk_{\bot}k_{\bot}\int_{-\infty}^{\infty}\!dk_{z}\;
\frac{e^{-k_{\bot}^2/\vb^2}e_n(k_{\bot}^2/\vb^2)}{k^2}\\
\times
\left[\frac{1}{\ve}+k_{\bot}^2F_n(k_{\bot},k_z)+(k^2+k_z^2)F_{n+1}(k_{\bot},k_z)\right],
\end{multline}
where
\be
F_n(k_{\bot},k_z)=\frac{1}{\ve[(\ve+k)^2-m^2]},
\ee
and $k=\sqrt{k_{\bot}^2+k_z^2}$.
Note that the function $\exp(-x)e_n(x)$ ($x=k_{\bot}^2/\vb^2$) is the Poisson distribution. 
When $x$ becomes large, the Poisson distribution is approximated as the Gaussian distribution 
centered at $n\sim x$ with its width $\sim1/\sqrt{x}$. 
Then, in the $\vb\to 0$ limit, $\exp(-x)e_n(x)$ has a sharp peak around $n\simeq k_{\bot}^2/\vb^2 $. 
On the other hand, 
the rest (terms inside a square bracket in eq.~(\ref{eshift2})) in the above integral is a slowly varying function of $n$. 
Therefore, the above integral is approximated by expanding a slowly varying function around the peak. 
Namely, let $f(n)$ be the slowly varying function of $n$, then the summation is approximated by 
\bal
\sum_{n=0}^{\infty}e^{-x}e_n(x)f(n)
&=\sum_{n=0}^{\infty}e^{-x}e_n(x) \sum_{j=0}^{\infty}[\frac{f^{(j)}(x)}{j!}(n-x)^j]\\ \label{resum1}
&=f(x)+xf^{(2)}(x)/2!+xf^{(3)}(x)/3!+(3x^2+x)f^{(4)}(x)/4!+\cdots,
\end{align}
where $\sum_{n=0}^{\infty}e_n(x)=\exp(x)$ is used.  In order to evaluate the last term in eq.~(\ref{eshift2}), 
the following different version of that summation formula will be used,  
\bal
\sum_{n=1}^{\infty}e^{-x}e_{n-1}(x)f(n)
&=\sum_{n=1}^{\infty}e^{-x}e_{n-1}(x)\sum_{j=0}^{\infty}[\frac{f^{(j)}(x)}{j!}(n-x)^j]\\ \nonumber 
&=f(x)+f^{(1)}(x)+(x+1) f^{(2)}(x)/2!+(4x+1)f^{(3)}(x)/3!\\ \label{resum2}
&\hspace{3cm}+(3x^2+11x+1)f^{(4)}(x)/4!+\cdots.
\end{align}
Since the magnetic moment is given by the term proportional to $\vb^2=-2eB$, 
it is enough to keep only those terms in the following. 
Using this resummation technique, the terms proportional to $\vb^2$ are obtained by  
\bal
\sum_{n=0}^{\infty}&e^{-x}e_n(x)\frac{1}{\ve}\to \frac {3\vb^2}{8} \frac{k_{\bot}^2}{(m^2+k^2)^{5/2}},\\
\sum_{n=0}^{\infty}&e^{-x}e_n(x)k_{\bot}^2F_n\\ \nonumber
&\to \frac {\vb^2}{16} \frac{k_{\bot}^4}{k^2(m^2+k^2)^{1/2}}
\left\{\frac{3}{(m^2+k^2)^{2}}+\frac{1}{k^2(m^2+k^2)}-\frac{1}{k^2[(m^2+k^2)^{1/2}+k]^2} \right\},
\end{align}
and 
\bal
\sum_{n=1}^{\infty}&e^{-x}e_{n-1}(x)(k^2+k_z^2)F_n \to -\frac{\vb^2}{4}\frac{k^2+k_z^2}{k^2(m^2+k^2)^{3/2}}\\
&+\frac {\vb^2}{16} \frac{k_{\bot}^2(k^2+k_z^2)}{k^2(m^2+k^2)^{1/2}}
\left\{\frac{3}{(m^2+k^2)^{2}}+\frac{1}{k^2(m^2+k^2)}-\frac{1}{k^2[(m^2+k^2)^{1/2}+k]^2} \right\}. \nonumber
\end{align}
After these resummation, the integral yields the terms proportional to $\vb^2$ as  
\begin{multline} \label{eshift3}
\Delta E_g^{(\vb^2)}=\frac{m\alpha\vb^2}{2\pi}\int_{0}^{\infty}\!dk_{\bot}\int_{-\infty}^{\infty}\!dk_{z}\;\frac{k_{\bot}}{k^2}
\left[ \frac{3k_{\bot}^2}{4(m^2+k^2)^{5/2}}-\frac{k^2+k_z^2}{4k^2(m^2+k^2)^{3/2}}\right.\\
+\left.\frac{k_{\bot}^2}{8k^2(m^2+k^2)^{3/2}} -\frac{k_{\bot}^2}{8k^2(m^2+k^2)^{1/2}[(m^2+k^2)^{1/2}+k]^2} \right].
\end{multline}
Lastly the following elementary integrals are carried out :   
\bal
\int_{0}^{\infty}\!dk_{\bot}\int_{-\infty}^{\infty}\!dk_{z}\;\frac{k_{\bot}^3}{k^2(m^2+k^2)^{5/2}}
&=\frac{4}{9m^2},\\
\int_{0}^{\infty}\!dk_{\bot}\int_{-\infty}^{\infty}\!dk_{z}\;\frac{k_{\bot}(k^2+k_z^2)}{k^4(m^2+k^2)^{3/2}}
&=\frac{8}{3m^2},\\ 
\int_{0}^{\infty}\!dk_{\bot}\int_{-\infty}^{\infty}\!dk_{z}\;\frac{k_{\bot}^3}{k^4(m^2+k^2)^{3/2}}&=\frac{4}{3m^2},
\end{align}
and 
\be
\int_{0}^{\infty}\!dk_{\bot}\int_{-\infty}^{\infty}\!dk_{z}\;\frac{k_{\bot}^3}{k^4(m^2+k^2)^{1/2}[(m^2+k^2)^{1/2}+k]^2}
=\frac {4}{3m^2} \int_{0}^{\infty}\!dx\;\frac{1}{(1+x)^2}=\frac{2}{3m^2},
\ee
where the integration variable is changed to $k/(m^2+k^2)^{1/2}=x$ in the last integral.
Combining these integrals, the final result for the energy shift in the weak magnetic field limit is 
\bal
\Delta E_g^{(\vb^2)}&=
\frac{\alpha\vb^2}{2\pi m}\left(\frac34\cdot\!\frac49-\frac14\cdot\!\frac83+\frac18\cdot\!\frac43-\frac18\cdot\!\frac23 \right)\\
&=-\frac{\alpha}{2\pi}\left(\frac{-eB}{2m}\right)\\
&=-\frac{\alpha}{2\pi}\mu_B B,
\end{align}
where $\mu_B=|e|\hbar/2mc$ is the Bohr magneton, and all other terms except those proportional to $\vb^2$ are neglected. 
The geomagnetic factor $g$ of the electron is 2 from the Dirac equation. 
If the true geomagnetic factor including radiative corrections is expressed as $g=2(1+a)$, then 
the interaction energy between the magnetic moment and external magnetic fields $B$ is 
$E=-(1+a)\mu_B B$. Hence the energy shift is written as $\Delta E=-a\mu_B B$. From the above result, $a$ 
is found in the first radiative correction as  
\be
a=\frac{g-2}{2}=+\frac{\alpha}{2\pi}.
\ee
This is precisely the result first obtained by Schwinger with the aid of subtractions 
to remove the divergence of integrals \cite{schwinger48,schwinger49}. 

From the above derivation, it now becomes clear that the physical meaning of the anomalous magnetic moment 
and the role of Landau levels to it.  For the weak external magnetic field case, separation of these Landau levels are 
sufficiently small such that electrons can be excited within a very short time and come back to the ground state 
by interacting with photons surrounding them. 
As we have seen, in quantum electrodynamics these virtual transitions happen equally even to the higher Landau levels. 
This is understood from the fact that the Poisson distribution is well approximated with the Gaussian distribution in 
the weak magnetic field limit. In this sense, existence of the anomalous magnetic moment is a proof of 
underlying infinite number of Landau levels, even if we do not  observe them directly. 

\section{Divergent Terms and Zero Magnetic Field Limit}
We now examine divergent terms appeared in the weak magnetic field limit of the ground state energy. 
These terms are independent of this applied external field, i.e., terms of order $\vb^0$. 
Using the resummation technics in the previous section, we obtain 
\bal
\Delta E_g^{(\vb^0)}&=\frac{m\alpha}{2\pi}\int_{0}^{\infty}\!dk_{\bot}\int_{-\infty}^{\infty}\!dk_{z}\;\frac{k_{\bot}}{k^2}
\left[\frac{1}{ \sqrt{m^2+k^2}}+\frac{k}{ \sqrt{m^2+k^2} ( \sqrt{m^2+k^2}+k)} \right]\\ \label{div1}
&=\frac{m\alpha}{2\pi}\;\left[ \frac{\sqrt{m^2+k^2}}{\sqrt{m^2+k^2}+k}
+\frac32 \ln\left(\frac{\sqrt{m^2+k^2}+k}{\sqrt{m^2+k^2}-k}\right) \right]^{k\to\infty}_{k\to 0}.
\end{align}
Obviously, the second term in eq. (\ref{div1}) contains ultraviolet divergent term. 
This divergent term has the form $3m\alpha/(4\pi) \ln(k_{{\rm max}}^2/m^2)$ $(k_{{\rm max}}\to\infty)$, 
which coincides with the same divergent behavior as the case of radiative 
correction to the electron self energy without external fields \cite{itzykson}. 
We remark that this divergent behavior is also clearly seen in the last term of the expression (\ref{eshiftj}). 

Therefore, the mass of the electron needs to be renormalized. 
In other words the bare mass $m$ appeared in this formalism is not a physical mass. 
Alternatively, the mass should be defined in the limit $B\to 0$ in the modified Dirac equation (\ref{exact1}) in Schwinger's language, 
since the physical mass is not $m$ but $\tilde{{\cal M}}$.  

We next evaluate terms proportional to $\vb^4$, i.e. next order in the weak magnetic field limit expansion. 
Using the formulae (\ref{resum1},\ref{resum2}), we obtain 
\bal
\sum_{n=0}^{\infty}e^{-x}e_n(x)f_1(n)&\to x f^{(3)}(x)/3!+3x^2f^{(4)}(x)/4!\\
\sum_{n=1}^{\infty}e^{-x}e_{n-1}(x)f_2(n)&\to f^{(2)}(x)/2!+ 4x f^{(3)}(x)/3!+3x^2f^{(4)}(x)/4! ,
\end{align}
where $f_1(n)$ is either $1/\ve$ or $k_{\bot}^2 F_n(k_{\bot},k_z)$, and $f_2(n)=(k^2+k_z^2) F_n(k_{\bot},k_z)$. 
Denoting $\tilde{k}=\sqrt{k^2+m^2}$, straightforward calculations lead to 
\bal
\frac{\pd^3}{\pd n^3}\left.\left( \frac{1}{\ve}\right)\right|_{n=k_{\bot}^2/\vb^2}&=
-\frac{15\vb^6}{8\tilde{k}^7},\\ 
\frac{\pd^4}{\pd n^4}\left.\left( \frac{1}{\ve}\right)\right|_{n=k_{\bot}^2/\vb^2}&=
\frac{105\vb^8}{16\tilde{k}^9}, 
\end{align}
and 
\bal
\left.\frac{\pd^2F_n}{\pd n^2}\right|_{n=k_{\bot}^2/\vb^2}&=
\frac{3\vb^4}{8k^2\tilde{k}^5}+\frac{\vb^4}{8k^4\tilde{k}^3}-\frac{\vb^4}{8k^4\tilde{k}(k+\tilde{k})^2},\\
\left.\frac{\pd^3F_n}{\pd n^3}\right|_{n=k_{\bot}^2/\vb^2}&=
-\frac{15\vb^6}{16k^2\tilde{k}^7}-\frac{3\vb^6}{16k^4\tilde{k}^5},\\ \nonumber
\left.\frac{\pd^4F_n}{\pd n^4}\right|_{n=k_{\bot}^2/\vb^2}&=
\frac{105\vb^8}{32k^2\tilde{k}^9}+\frac{45\vb^8}{32k^4\tilde{k}^7}+\frac{9\vb^8}{32k^6\tilde{k}^5}\\
\hspace{2cm}-&\frac{3\vb^8}{32k^8\tilde{k}^3}+\frac{3\vb^8}{32k^8\tilde{k}(k+\tilde{k})^2} +\frac{3\vb^8}{16k^7\tilde{k}(k+\tilde{k})^3}.
\end{align}
Careful evaluations for the above integrals show that there are also divergent terms appeared. 
To identify leading behavior of divergence, we keep finite boundaries for integrals. 
The final expression is then, 
\bal 
\Delta E_g^{(\vb^4)}&=\frac{m\alpha\vb^4}{2\pi} \left[\frac{181}{64}I^{(1)} 
-\frac{475}{672}I^{(2)} +\frac{41}{420}I^{(3)}-\frac{293}{16800}I^{(4)}+\frac{863}{16800}I^{(5)} \right], 
\end{align}
where we have defined the integrals $I^{(i)}$ ($i=1\sim5$) as follows. 
\bal
I^{(1)}&=\int_{k_{{\rm min}}}^{k_{\rm max}}\!dk\; \frac{k^4}{\tilde{k}^9}, \\
I^{(2)}&=\int_{k_{{\rm min}}}^{k_{\rm max}}\!dk\; \frac{k^2}{\tilde{k}^7}, \\
I^{(3)}&=\int_{k_{{\rm min}}}^{k_{\rm max}}\!dk\; \frac{1}{\tilde{k}^5}, \\
I^{(4)}&=\int_{k_{{\rm min}}}^{k_{\rm max}}\!dk\; \left(-\frac{1}{k^2 \tilde{k}^3}
+\frac{1}{k^2\tilde{k}(k+\tilde{k})^2}+\frac{2} {k\tilde{k}(k+\tilde{k})^3}\right), \\
I^{(5)}&=\int_{k_{{\rm min}}}^{k_{\rm max}}\!dk\; \frac{1}{k\tilde{k}(k+\tilde{k})^3}.
\end{align}
It is easy to see to observe that all integrals except $I^{(5)}$ have finite values 
in the limit $k_{{\rm min}}\to 0$ and $k_{{\rm max}}\to\infty$. 
Those values are 
\be
I^{(1)}=\frac{2}{35m^4},\;I^{(2)}=\frac{2}{15m^4},\;I^{(3)}=\frac{2}{3m^4},\;I^{(4)}=-\frac{1}{m^4}, 
\ee
and $I^{(5)}$ is 
\be
I^{(5)}=\frac{1}{m^4}\left[ \frac{2\sqrt{k^2+m^2}}{k+\sqrt{k^2+m^2}}+\ln\left( \frac{k}{k+\sqrt{k^2+m^2}}\right)
\right]^{k_{\rm max}}_{k_{\rm min}}.
\ee
Thus, infinity arises in the limit $k_{\rm min}\to 0$ as $\ln(k_{\rm min}/m)$. 
However, we notice that this infinity is originated from improper use of the resummation 
formulae (\ref{resum1},\ref{resum2}). Indeed, these formulae fail for very small $k$, 
where the function $f(n)$ is no longer slowly varying function of $n$. 
The same difficulty is also encountered when one tries to evaluate 
the weak magnetic field limit using the proper time representation eq.~(\ref{eshiftsch}) \cite{psf}.
However, we remark that the anomalous magnetic moment of an electron can be 
obtained in the weak field limit without any divergence by employing the definition (\ref{defmu}).

\section{Lowest Landau Level and Strong Magnetic Field Limit}
In this section we utilize the lowest Landau level approximation in the strong magnetic field limit. 
For sufficient large values of $B$, energy levels are separated by large enough such that 
higher Landau levels will not contribute to the radiative correction compared to the lowest Landau level. 
In this approximation we obtain the leading term from the expression (\ref{eshiftj}) as 
\be
\Delta E_g^{{\rm LLL}}=\frac{m\alpha}{2\pi} \int_{0}^{\infty}\!d\omega\;2 e^{-\omega/\eta} I(\omega;0) . 
\ee
It is convenient to change integration variables to get 
\be
\Delta E_g^{{\rm LLL}}=\frac{m\alpha}{2\pi}e^{-2/\eta} \left( \int_0^{\pi}\!dz\;z\;e^{-(2 \cos z)/\eta}
+\int_0^{\infty}\!dz\;z\;e^{-(2 \cosh z)/\eta} \right). 
\ee
This integral representation shows that, in this approximation, there is no singular behavior in the strong magnetic field limit. 
Further corrections can be obtained by taking into account contributions from the higher Landau levels. 

\section{Conclusion and Discussion}
We have shown that a proper choice of the Fock space would not lead to divergent integrals when obtaining 
the radiative correction to the anomalous magnetic moment of the electron. 
However, $m$ and $e$ appeared in this formalism are not physical observed mass and charge respectively. 
These quantities should be defined self-consistently. 

Another issue which we have not discussed in details concerns applications to 
strong external field problems in quantum electrodynamics. 
There have been many activities on this issue in the past. 
Essentially the same quantization scheme is used to deal with 
intense external fields problems. From our discussions the reason is clear  
why we need to quantize electron fields in classical external fields. 
If one starts from a Fock vacuum for non-interacting electrons, as is done 
in ordinary quantum electrodynamics, then one {\it cannot} obtain 
the true vacuum for interacting electrons in strong fields in a general perturbative manners. 
Therefore it is crucial to choose the initial Fock vacuum as close as possible the real interacting electrons in external fields. 

In this paper we have also derived the alternative expression for the energy shift of the ground state 
in a uniform magnetic field. Detailed analysis of our expression will be studied together with applications to 
strong magnetic field separately \cite{jun}. These are especially relevant to recent astrophysical observations.  

In conclusion, the renormalization procedure is necessary in current quantum field theories. 
since we do not know how to solve interacting systems in general. 
The structures of quantum field theories themselves, however, seem to have less to do with the 
reason for divergent numbers. Rather, the problem is originated from 
the simplifications in our theoretical models, such as an assumption of non-interacting particles, 
infinite volumes, infinite number of particles, and so on.  
We then have a ground to conclude that a better choice of the Fock vacua may not 
necessarily result in divergence difficulties. 

\section*{acknowledgments}
The author would like to express his acknowledgment to Professor Pawel O. Mazur for 
suggesting this problem and many valuable discussions. 
He would also like to thank Andreas Keil for constructive discussions. 
This work was supported in part by the National Science Foundation, Grant No. PHY-0140377, 
to the University of South Carolina, and by the National University of Singapore. 



\begin{thebibliography}{99}
\bibitem{comment1} This quantization method is also known as ``Furry's representation" 
which was introduced in Ref.~\citen{furry}.
\bibitem{comment2} Ritus's work are reviewed in Ref.~\citen{ritus}. See also Ref.~\citen{psf}.
\end{thebibliography}
\end{document}